\documentclass[10pt,conference]{IEEEtran}
\IEEEoverridecommandlockouts
\usepackage{bm}
\usepackage{makeidx}
\usepackage{enumerate}
\usepackage{color}
\usepackage{cite}
\usepackage{amsmath,amsthm}   
\usepackage{amssymb}
\usepackage{multirow}
\usepackage{graphicx}
\usepackage{epstopdf}
\usepackage{balance}
\usepackage{multicol}
\DeclareGraphicsExtensions{.pdf,.jpeg,.png,.jpg,.emf,.eps}
\definecolor{darkgreen}{RGB}{0, 122, 0}
\definecolor{darkred}{RGB}{180, 0, 0}
\definecolor{darkblue}{rgb}{0,0.41,0.7}

\usepackage{calc}
\usepackage{url}

\usepackage{upref}
\usepackage{comment}
\usepackage{times}
\usepackage{dsfont}
\usepackage{rawfonts}
\usepackage[T1]{fontenc}
\usepackage{latexsym}
\usepackage{amsfonts}

\usepackage{fontawesome}

\usepackage{optidef}
\usepackage[compress]{cleveref}
\usepackage{lipsum}

\usepackage{xcolor}
\usepackage{tikz,ifthen,calc,graphicx,pgfplots}
\pgfplotsset{compat=1.18}
\usetikzlibrary{plotmarks}
\usetikzlibrary{arrows.meta}
\usetikzlibrary{fit}
\usetikzlibrary{decorations.pathreplacing}

\usetikzlibrary{calc}
\usetikzlibrary{intersections}
\usetikzlibrary{arrows,shapes}
\usetikzlibrary{positioning}
\usetikzlibrary{plotmarks}
\pgfplotsset{compat=1.18}

\theoremstyle{definition}

\begin{document}

\title{Low Probability of Detection Communication Using Noncoherent Grassmannian Signaling\\
\thanks{This work was supported under grant PID2022-137099NB-C43 (MADDIE) funded by MICIU/AEI /10.13039/501100011033 and FEDER, UE. The work of Diego Cuevas was also partly supported under grant FPU20/03563 funded by MIU, Spain.}}

\author{%
    \IEEEauthorblockN{Diego Cuevas, Mikel Guti{\'e}rrez, Jes{\'u}s Ib{\'a}ñez and Ignacio Santamaria}
    \vspace{0.20cm}
    \IEEEauthorblockA{Dept. of Communications Engineering \\ Universidad de Cantabria, Spain \\ email: diego.cuevas@unican.es, mikel.gutierrez@alumnos.unican.es, \\ jesus.ibanez@unican.es, i.santamaria@unican.es}
}

\maketitle

\begin{abstract}

This paper proposes a noncoherent low probability of detection (LPD) communication system based on direct sequence spread spectrum (DSSS) and Grassmannian signaling. Grassmannian constellations enhance covertness because they tend to follow a noise-like distribution. Simulations showed that Grassmannian signaling provides competitive bit error rates (BER) at low signal-to-noise ratio (SNR) regimes with low probability of detection at the unintended receiver compared to coherent schemes that use QPSK or QAM modulation formats and need pilots to perform channel estimation. The results suggest the practicality and security benefits of noncoherent Grassmannian signaling for LPD communications due to their improved covertness and performance.

\end{abstract}

\IEEEpeerreviewmaketitle

\section{Introduction}

In various scenarios where communication security is critical, the bare detection of a transmission can disclose significant information (e.g., the location of a transmitter). Therefore, it is essential not only to ensure the security of the transmitted data but also to protect the transmission itself \cite{yan_low_2019}. In an LPD system, a trusted transmitter (Alice) attempts to communicate with a trusted receiver (Bob) while a warden (Willie) monitors the channel, trying to detect the communication \cite{bash_hiding_2015} (cf. Fig. \ref{fig:SystemModel}). Under the hypothesis that Alice is not transmitting, Willie acquires additive white Gaussian noise (AWGN). It has been demonstrated in \cite{yan_gaussian_2019} that Gaussian signaling is optimal when $\mathbb{D}(p_{1}||p_{0}) \leq 2 \epsilon^2$ is used as a covertness constraint, where $\mathbb{D}(\cdot||\cdot)$ denotes the Kullback-Leibler (KL) divergence and $p_{1}$ and $p_{0}$ are the distributions of the signal received by Willie when Alice is or is not transmitting, respectively. It seems intuitive that minimizing the KL divergence between the noise distribution and that of the communications signal transmitted by Alice enhances the undetectability of the communication.

Typically, LPD communications use coherent signaling schemes where it is common to find the assumption of perfect channel state information (CSI) at Bob's receiver \cite{okamoto_chaos_2011,zheng_multi-antenna_2019}. In practice, obtaining accurate CSI requires sending a sufficiently large number of pilots. However, the insertion of pilots within the transmitted frame introduces periodicities or other sources of cyclostationarity in the signal, thus facilitating Willie's detection of the transmission \cite{yan_low_2019}. It is therefore of interest to study noncoherent signaling schemes for LPD communications that avoid the use of pilots. We assume that Alice, Bob, and Willie are each equipped with a single antenna. This assumption is reasonable as LPD networks are sometimes deployed by dismounted crew, as in mobile ad hoc networks (MANETs) \cite{regragui_dynamics_2018} and thus the terminals need to be compact and lightweight.

To the best of the authors' knowledge, only the recent work in \cite{katsuki_new_2023} has considered a noncoherent scheme for LPD communications. However, this work is based on differential modulations and therefore the transmission of a reference pilot or space-time codeword per frame between Alice and Bob is required. To avoid sending pilots or any other reference signal, in this paper we propose a noncoherent scheme based on Grassmannian or unitary space-time constellations \cite{hochwald_unitary_2000,lizhong_zheng_communication_2002,Dhillon2008constructing,gohary,Ngo_cubesplit_journal}, which are the optimal signals from an information-theoretic point of view for noncoherent communications.

The proposed LPD scheme is a direct sequence spread spectrum (DSSS) system modulated with Grassmannian codewords, which are designed using the manifold optimization algorithm proposed in \cite{cuevas_fast_2021,cuevas_union_2023}. The DSSS system brings the signal below the noise level, making it difficult for Willie to detect the transmission using energy detectors, while the use of the noncoherent Grassmanian constellation generates noise-like signals that are more difficult for Willie to detect using Gaussianity detectors. We compare the bit error rate (BER) of the proposed DSSS system modulated with Grassmannian symbols against various traditional quadrature phase shift keying (QPSK) and quadrature amplitude modulations (QAM), evaluating their detectability through numerical simulations. Our simulation results show that the noncoherent Grassmannian system provides a competitive BER at low signal-to-noise ratio (SNR) regime, while also generating noise-like signals, which enhances its undetectability.

\section{System Model}

\begin{figure}[t!]
     \centering
     \input{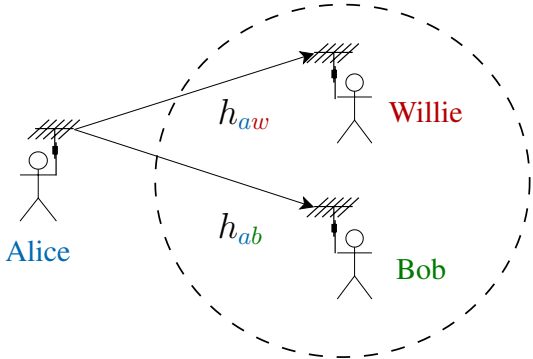}
     \vspace{-.5cm}
     \caption{Illustration of the system model for LPD communications.}
     \label{fig:SystemModel}
 \end{figure}

Fig. \ref{fig:SystemModel} shows the LPD communications scenario considered in this work, where Alice, Bob, and Willie each have a single antenna. We assume that the channel between Alice and Bob ($h_{ab}$) and between Alice and Willie ($h_{aw}$) is a frequency-flat Rayleigh block-fading channel, which follows a complex normal distribution with mean 0 and variance 1 (i.e., $h_{ab}, h_{aw} \sim \mathcal{CN} \left(0, 1\right)$). The channels remain constant during each coherence block of $T$ symbols and change to an independent realization in the next block, both for Bob and Willie. The channels are unknown to all transceivers. In each coherence block, the transmitter sends a codeword $\mathbf{x}[i]$ chosen uniformly from the codebook $\mathcal{C}=\{\mathbf{x}_1,\dotsc,\mathbf{x}_K\} \subset \mathbb{C}^T$, where each codeword carries $\log_{2}(K)$ bits of information.

At the $i$-th coherence block, the signal received by Bob $\mathbf{y}_{b}[i]\in \mathbb{C}^{T}$ is

\begin{equation}
\mathbf{y}_{b}[i] = \mathbf{x}[i] h_{ab} + \mathbf{n}_{b}[i]\, ,
\end{equation}
where $\mathbf{n}_{b}[i]$ is additive white Gaussian noise (AWGN) modeled by a complex normal distribution of mean 0 and variance $\sigma^{2}_{b}$ (i.e., $\mathbf{n}_b[i] \sim \mathcal{CN} \left(0, \sigma^{2}_{b}\right)$). Likewise, the signal received by Willie is 

\begin{equation}
\mathbf{y}_{w}[i] = \mathbf{x}[i] h_{aw} + \mathbf{n}_{w}[i]\, ,
\end{equation}where $\mathbf{n}_{w}[i]$ is AWGN modeled by a complex normal distribution of mean 0 and variance $\sigma^{2}_{w}$ (i.e., $\mathbf{n}_w[i] \sim \mathcal{CN} \left(0, \sigma^{2}_{w}\right)$).

\section{Proposed Grassmannian LPD Scheme}\label{sec:proposed_scheme}

\begin{figure}[t!]
    \centering
    \input{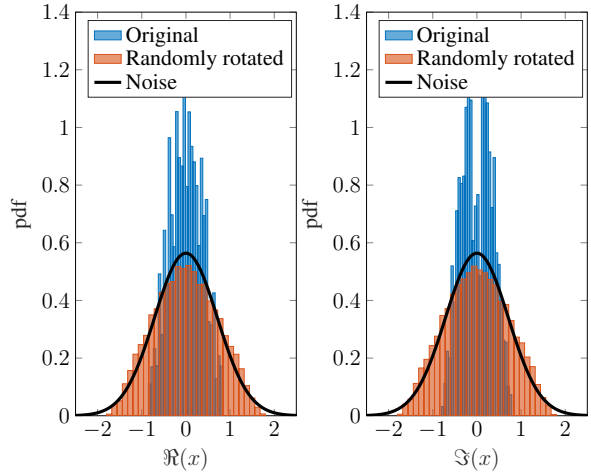}
    \vspace{-.5cm}
    \caption{Histograms of the real and imaginary parts of the transmitted Grassmannian codewords with and without random phase rotations for $T = 4$ time slots and $K = 64$ codewords.}
    \label{fig:Histograms}
\end{figure}

In this work, a noncoherent communications scheme that uses Grassmannian constellations for LPD communications is proposed. The codewords $\mathbf{x}_k \in \mathcal{C}$ are unitary vectors, $\mathbf{x}^H_k\mathbf{x}_k=1$, representing $K$ subspaces from the
Grassmann manifold of 1-dimensional subspaces on $\mathbb{C}^T$, denoted here as $\mathbb{G}(1,\mathbb{C}^T)$ (for an introduction to the Grassmann manifold, see \cite[Ch. 9]{Coherence}). The constellation is designed with a gradient ascent algorithm that operates directly on the Grassmann manifold to maximize the minimum chordal distance between codewords \cite{cuevas_fast_2021}, which is a commonly used metric in noncoherent communications \cite{edelman_geometry_1998}. By definition of the Grassmann manifold, the codeword $\mathbf{x}_k$ and its rotated version $\mathbf{x}_ke^{j\theta}$ represent the same point in $\mathbb{G}(1,\mathbb{C}^T)$. This property allows us to transmit randomly rotated versions of the codewords, thereby achieving a noise-like constellation. This can be seen in Fig. \ref{fig:Histograms}, where we illustrate the effect of this random rotation on the distribution of the transmitted signals. The histograms show the distribution of the real and imaginary parts of the original Grassmannian codewords and their randomly rotated versions. As it can be observed, after applying random phase rotations, the histograms resemble Gaussian-like noise rather than a deterministic constellation. This confirms that the rotation operation effectively masks the underlying codeword structure, producing a noise-like signal while preserving the fundamental invariance of the Grassmann manifold.

We have considered a relatively small coherence block of $T$ = 4 time slots. This choice avoids introducing excessive delay in the communication and ensures the system operates effectively even in ultra-high mobility environments (e.g. communications between aircraft or air-to-ground communications). The size of the Grassmannian constellation is $K$ = 64 codewords, which results in a spectral efficiency of $\eta=\frac{\log_{2}(K)}{T}$ = 1.5 bits/s/Hz. As previously mentioned, the system is single antenna, although it could be easily extended to a multiple-input multiple-output (MIMO) system since the optimization algorithm in \cite{cuevas_fast_2021,cuevas_union_2023} allows the design of Grassmannian MIMO constellations. It is worth mentioning that the design of these constellations can be optimized for increased robustness against hardware impairments such as I/Q imbalance or carrier frequency offset \cite{cuevas_hardware_2024}. The codebook is known to Alice and Bob.

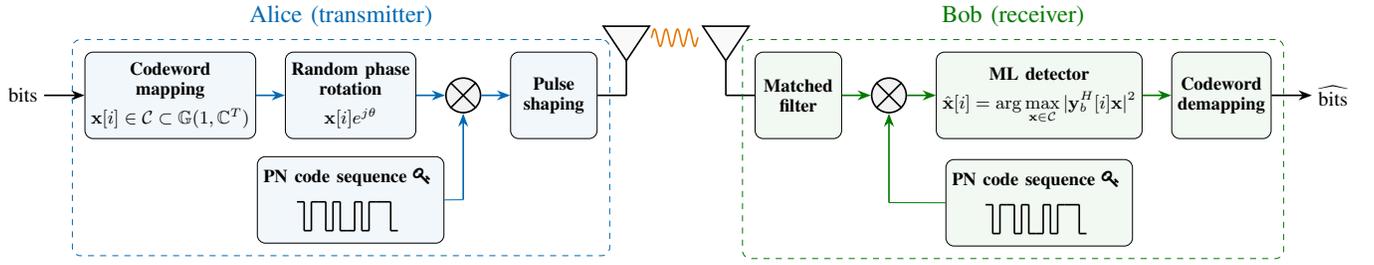
\begin{figure*}[t!]
\centering
\small
\definecolor{darkgreen}{RGB}{0, 122, 0}
\definecolor{darkred}{RGB}{180, 0, 0}
\definecolor{darkblue}{rgb}{0,0.41,0.7}
\definecolor{darkorange}{RGB}{230,120,0}

\resizebox{\textwidth}{!}{
\begin{tikzpicture}[
  block/.style = {draw, thick, rounded corners, minimum width=2.8cm, minimum height=1cm, align=center, fill=white},
  smallblock/.style = {draw, thin, rounded corners, minimum width=1.5cm, minimum height=1.5cm, align=center, fill=darkblue!5},
  circleblock/.style = {draw, thick, circle, minimum size=.6cm, align=center, fill=gray!10},
  arrow/.style = {->, >=Stealth, thick},
  arrowTX/.style = {->, >=Stealth, thick, draw=darkblue},
  arrowRX/.style = {->, >=Stealth, thick, draw=darkgreen},
  annot/.style = {font=\footnotesize, align=center},
  antenna/.style = {thick}
  ]

\node[smallblock] (bit) {\bfseries Codeword\\ \bfseries mapping \\[.15cm] $\mathbf{x}[i] \in\mathcal{C}\subset\mathbb{G}(1,\mathbb{C}^T)$};
\node[smallblock, right=5mm of bit] (rot) {\bfseries Random phase\\ \bfseries rotation\\[.15cm] $\mathbf{x}[i] e^{j\theta}$};
\node[circleblock, right=5mm of rot] (pn) {};
\node[smallblock, right=5mm of pn] (shape) {\bfseries Pulse\\\bfseries shaping};

\node[smallblock, below=3mm of rot] (pngen) {
  \shortstack{{\bfseries PN code sequence} \faKey\\[.15cm]
  \tikz[scale=0.5,baseline={(0,-0.25)}]{
    \draw[thick] (0,.5) -- (.25,.5);                  
    \draw[thick] (.25,.5) -- (.25,-.5);     
    \draw[thick] (.25,-.5) -- (.5,-.5);
    \draw[thick] (.5,-.5) -- (.5,.5);         
    \draw[thick] (.5,.5) -- (.75,.5);
    \draw[thick] (.75,.5) -- (1,.5);                  
    \draw[thick] (1,.5) -- (1,-.5);      
    \draw[thick] (1,-.5) -- (1.25,-.5);
    \draw[thick] (1.25,-.5) -- (1.25,.5);
    \draw[thick] (1.25,.5) -- (1.5,.5);
    \draw[thick] (1.5,.5) -- (1.5,-.5);
    \draw[thick] (1.5,-.5) -- (1.75,-.5);
    \draw[thick] (1.75,-.5) -- (2,-.5);
    \draw[thick] (2,-.5) -- (2,.5);
    \draw[thick] (2,.5) -- (2.25,.5);
    \draw[thick] (2.25,.5) -- (2.25,-.5);
    \draw[thick] (2.25,-.5) -- (2.5,-.5);
    \draw[thick] (2.5,-.5) -- (2.5,.5);
    \draw[thick] (2.5,.5) -- (2.75,.5);
    \draw[thick] (2.75,.5) -- (3,.5);
    \draw[thick] (3,.5) -- (3.25,.5);
    \draw[thick] (3.25,.5) -- (3.25,-.5);
    \draw[thick] (3.25,-.5) -- (3.5,-.5);
  }}
};

\draw[thick] (pn.north west) -- (pn.south east);
\draw[thick] (pn.north east) -- (pn.south west);

\node[below=13.85mm of pn] (linemul) {};
\draw[draw=darkblue] (pngen.east) -- (linemul.center);
\draw[arrowTX] (linemul.center) -- (pn.south);

\node[right=5mm of shape] (txant) {};
\draw[antenna] (shape.east) -- (txant); 
\draw[antenna] (txant.west) -- ++(0,0.6); 
\draw[antenna, fill=gray!5] (txant.west) ++(0,0.6) -- ++(-0.4,0.6) -- ++(0.8,0) -- cycle; 

\node[smallblock, fill=darkgreen!5, right=20mm of txant] (rx) {\bfseries Matched\\\bfseries filter};
\node[circleblock, right=5mm of rx] (dspread) {};
\node[smallblock, fill=darkgreen!5, right=5mm of dspread] (ml) {\bfseries ML detector\\[.15cm]$\displaystyle \hat{\mathbf{x}}[i]=\arg\max_{\mathbf{x}\in\mathcal{C}}|\mathbf{y}_b^H[i]\mathbf{x}|^{2}$};
\node[smallblock, fill=darkgreen!5, right=5mm of ml] (out) {\bfseries Codeword\\\bfseries demapping};


\node[smallblock, fill=darkgreen!5, below=3.5mm of ml] (pngenrx) {
  \shortstack{{\bfseries PN code sequence} \faKey\\[.15cm]
  \tikz[scale=0.5,baseline={(0,-0.25)}]{
    \draw[thick] (0,.5) -- (.25,.5);                  
    \draw[thick] (.25,.5) -- (.25,-.5);     
    \draw[thick] (.25,-.5) -- (.5,-.5);
    \draw[thick] (.5,-.5) -- (.5,.5);         
    \draw[thick] (.5,.5) -- (.75,.5);
    \draw[thick] (.75,.5) -- (1,.5);                  
    \draw[thick] (1,.5) -- (1,-.5);      
    \draw[thick] (1,-.5) -- (1.25,-.5);
    \draw[thick] (1.25,-.5) -- (1.25,.5);
    \draw[thick] (1.25,.5) -- (1.5,.5);
    \draw[thick] (1.5,.5) -- (1.5,-.5);
    \draw[thick] (1.5,-.5) -- (1.75,-.5);
    \draw[thick] (1.75,-.5) -- (2,-.5);
    \draw[thick] (2,-.5) -- (2,.5);
    \draw[thick] (2,.5) -- (2.25,.5);
    \draw[thick] (2.25,.5) -- (2.25,-.5);
    \draw[thick] (2.25,-.5) -- (2.5,-.5);
    \draw[thick] (2.5,-.5) -- (2.5,.5);
    \draw[thick] (2.5,.5) -- (2.75,.5);
    \draw[thick] (2.75,.5) -- (3,.5);
    \draw[thick] (3,.5) -- (3.25,.5);
    \draw[thick] (3.25,.5) -- (3.25,-.5);
    \draw[thick] (3.25,-.5) -- (3.5,-.5);
  }}
};

\node[below=14.35mm of dspread] (linemulrx) {};
\draw[draw=darkgreen] (pngenrx.west) -- (linemulrx.center);
\draw[arrowRX] (linemulrx.center) -- (dspread.south);

\draw[thick] (dspread.north west) -- (dspread.south east);
\draw[thick] (dspread.north east) -- (dspread.south west);

\node[left=5mm of rx] (rxant) {};
\draw[antenna] (rxant) -- (rx.west); 
\draw[antenna] (rxant.east) -- ++(0,0.6); 
\draw[antenna, fill=gray!5] (rxant.east) ++(0,0.6) -- ++(-0.4,0.6) -- ++(0.8,0) -- cycle; 

\draw[thick,decorate,decoration={snake,amplitude=1.5mm,segment length=2mm}, draw=darkorange] 
     ($(txant.east)+(0.1,0.5)+(0.1,0.5)$) -- ($(rxant.west)+(-0.1,0.5)+(-0.15,0.5)$);

\node[left=7mm of bit] (bits) {\normalsize bits};
\draw[arrow] (bits) -- (bit.west);
\draw[arrowTX] (bit) -- (rot);
\draw[arrowTX] (rot) -- (pn);
\draw[arrowTX] (pn) -- (shape);

\draw[arrowRX] (rx) -- (dspread);
\draw[arrowRX] (dspread) -- (ml);
\draw[arrowRX] (ml) -- (out);
\node[right=7mm of out] (bitsout) {\normalsize $\widehat{\textnormal{bits}}$};
\draw[arrow] (out.east) -- (bitsout);

\node[draw, color=darkblue, dashed, rounded corners, fit={(bit) (rot) (pn) (shape) (pngen)}, inner sep=6pt] (TXbox) {};
\node[annot, above=1mm of TXbox.north] {\large \textcolor{darkblue}{Alice (transmitter)}};

\node[draw, color=darkgreen, dashed, rounded corners, fit={(rx) (dspread) (ml) (out) (pngenrx)}, inner sep=6pt] (RXbox) {};
\node[annot, above=1mm of RXbox.north] {\large \textcolor{darkgreen}{Bob (receiver)}};

\end{tikzpicture}
}
\vspace{-.25cm}
\caption{Block diagram of the proposed noncoherent Grassmannian LPD scheme.}
\label{fig:tx_rx_scheme}
\end{figure*}

Each symbol is multiplied by a secret pseudo-random noise (PN) code sequence, which consists of 31 chips and is known to Alice and Bob, followed by signal shaping. The use of DSSS not only allows us to spread the signal power below the noise floor, but also aids in achieving temporal synchronization. Otherwise, a noncoherent system would need some pilot signal to perform time synchronization. Thanks to the good autocorrelation properties of the DSSS spreading codes, however, 
the system can be synchronized without sending pilot signals.

At the receiver side, Bob applies the optimal noncoherent Maximum Likelihood (ML) detector for Grassmannian constellations, which is given by
\begin{equation}
\mathbf{\hat{x}}[i] = \arg \max_{\mathbf{x}\in \mathcal{C}}  |\mathbf{y}_b^H[i]\mathbf{x}|^{2}\, .
\end{equation}

A block diagram of the proposed LPD scheme could be seen in Fig. \ref{fig:tx_rx_scheme}.

\section{Binary Hypothesis Testing at Willie}
The goal of the warden, Willie, is to deliver a decision in each sensing period: $\mathcal{D}_{0}$ if no communication has been detected, and $\mathcal{D}_{1}$ if communication has been detected. For this purpose, Willie solves the following binary hypothesis testing problem: 
\begin{equation}
\left\lbrace\begin{array}{l}
\mathcal{H}_{0}: \mathbf{y}_{w}[i] = \mathbf{n}_{w}[i]  \\ 
\mathcal{H}_{1}: \mathbf{y}_{w}[i] = \mathbf{x}[i]h_{aw}+\mathbf{n}_{w}[i]\, ,
\end{array}\right.
\end{equation} 
where $\mathcal{H}_{0}$ is the null hypothesis and $\mathcal{H}_{1}$ is  the alternative hypothesis. The probability of false alarm is $\Pr(\mathcal{D}_{1}|\mathcal{H}_{0})$ and the probability of detection is $\Pr(\mathcal{D}_{1}|\mathcal{H}_{1})$, and the covertness constraint is typically set as a function of the probability of error at Willie.

One possibility for Willie is to employ an energy detector such as a radiometer \cite{lee_achieving_2015}. The performance of this detector depends essentially on the length of the spreading sequence and is the same for coherent and noncoherent systems. Therefore, in this work, we focus on Gaussianity tests that measure how far the received signal departs from the Gaussianity assumed under the $\mathcal{H}_{0}$ hypothesis.

We assume that the mean and variance of the noise under $\mathcal{H}_{0}$ at Willie's receiver are unknown. The performance of various Gaussianity tests has been evaluated: Jarque-Bera \cite{jarque_test_1987}, Shapiro-Wilk \cite{shapiro_analysis_1965}, Lilliefors \cite{lilliefors_h_w_kolmogorov-smirnov_nodate}, Chi-Squared \cite{spurr_goodness--fit_1988}, and D'Agostino-Pearson \cite{dagostino_goodness--fit-techniques_1986}. Based on their performance, we selected the Jarque-Bera test for Willie's detector, as it achieves a good balance between performance and computational cost. The Jarque-Bera test is a test with two degrees of freedom that measures kurtosis and skewness \cite{jarque_test_1987}. In our implementation, the test is conducted on signals sampled at the chip sampling rate and of length 50000 samples.


\section{Results}

In this section, we evaluate the performance of the proposed noncoherent LPD communication system through simulations and OTA experimental measurements. As a benchmark, we employ a coherent LPD communications system, in which the frames are modulated using QPSK or 64-QAM symbols. The frame's structure includes both information symbols and pilots, which are generated randomly and are normalized to have unit power. At Bob's receiver, the pilots are used for CSI estimation and the channel estimate thus obtained is then used by the linear minimum mean squared error (LMMSE) decoder.

In order to perform a fair comparison, all frames carry the same amount of information, ensuring they operate at the same spectral efficiency of $\eta$ = 1.5 bits/s/Hz. Thus, the QPSK frame consists of $T$-1 information symbols and 1 pilot, whereas the 64-QAM frame comprises 3 pilot symbols and 1 information symbol per coherence block. All frames have the same average power and have been spread with the same PN sequence and shaped in the same manner.

\begin{figure}[t!]
    \centering
%
%
\definecolor{mycolor1}{rgb}{0.00000,0.44700,0.74100}%
\definecolor{mycolor2}{rgb}{0.85000,0.32500,0.09800}%
\definecolor{mycolor3}{rgb}{0.92900,0.69400,0.12500}%
\definecolor{mycolor4}{rgb}{0.49400,0.18400,0.55600}%
\resizebox{\columnwidth}{!}{

\begin{tikzpicture}

\begin{axis}[%
width=4.521in,
height=3.566in,
at={(0.758in,0.481in)},
scale only axis,
font = \Large,
xmin=2,
xmax=8,
xtick={2,3,4,5,6,7,8},
xlabel style={font=\Large\color{white!15!black}},
xlabel={$T$},
ymode=log,
ymin=0.01,
ymax=1.29641013764055,
yminorticks=true,
ylabel style={font=\Large\color{white!15!black}},
ylabel={$D_{KL}\left(p(x) \,||\, \mathcal{CN}(0,1)\right)$},
axis background/.style={fill=white},
xmajorgrids,
ymajorgrids,
yminorgrids,
minor grid style = dashed,
legend style={legend cell align=left, align=left, draw=white!15!black}
]
\addplot [color=blue, line width=1.5pt, mark size=4.0pt, mark=square, mark options={solid, blue}]
  table[row sep=crcr]{%
2	1.29641013764055\\
4	0.490744738385613\\
6	0.208216251286433\\
8	0.109415903268989\\
};
\addlegendentry{$\eta\text{ = 0.5}$ bits/s/Hz}

\addplot [color=red, line width=1.5pt, mark size=5.0pt, mark=triangle, mark options={solid, red}]
  table[row sep=crcr]{%
2	0.742161874513865\\
4	0.164930630854746\\
6	0.0481442834773589\\
8	0.0266293823744957\\
};
\addlegendentry{$\eta\text{ = 1}$ bits/s/Hz}

\addplot [color=green, line width=1.5pt, mark size=4.0pt, mark=o, mark options={solid, green}]
  table[row sep=crcr]{%
2	0.484519656144792\\
4	0.0908093352619303\\
6	0.0312032572934735\\
8	0.019215999216556\\
};
\addlegendentry{$\eta\text{ = 1.5}$ bits/s/Hz}

\addplot [color=black, line width=1.5pt, mark size=5.0pt, mark=diamond, mark options={solid, black}]
  table[row sep=crcr]{%
2	0.380765279752708\\
4	0.0630038694805888\\
6	0.023202980217024\\
8	0.015619102628164\\
};
\addlegendentry{$\eta\text{ = 2}$ bits/s/Hz}

\end{axis}

\begin{axis}[%
width=5.833in,
height=4.375in,
at={(0in,0in)},
scale only axis,
xmin=0,
xmax=1,
ymin=0,
ymax=1,
axis line style={draw=none},
ticks=none,
axis x line*=bottom,
axis y line*=left
]
\end{axis}
\end{tikzpicture}%

}
    \vspace{-.5cm}
    \caption{Estimated KL divergence between the noise distribution and that of transmitted Grassmannian constellations designed with \cite{cuevas_fast_2021} as a function of the coherence time $T$ and the spectral efficiency $\eta$.}
    \label{fig:KL_divergence}
\end{figure}
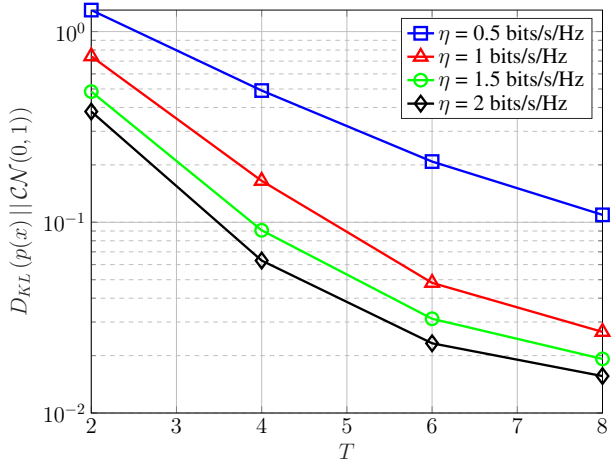

Before analyzing BER and detection probability, we first assess the statistical similarity between the transmitted Grassmannian constellations and Gaussian noise. This is quantified using the KL divergence, which measures the deviation of the constellation distribution from the noise distribution. Intuitively, minimizing the KL divergence directly improves the undetectability of the communication, since the signal becomes statistically harder to distinguish from background noise. Fig. \ref{fig:KL_divergence} shows the estimated KL divergence (computed using the method described in \cite{PerezCruz2008}) between the noise distribution and that of Grassmannian constellations, evaluated for different values of coherence time $T$ and spectral efficiency $\eta$. As expected, the KL divergence decreases with increasing coherence time, since longer codewords spread the signal energy more evenly and make the transmitted distribution more Gaussian-like. Similarly, for fixed $T$, higher spectral efficiency (larger constellations) also reduces the divergence, as the greater number of codewords produces a distribution that more closely approximates Gaussian noise.

As explained in Section \ref{sec:proposed_scheme}, in our work we focus on the operating point $T = 4$ and $\eta = 1.5$ bits/s/Hz, where the use of a short coherence block length preserves robustness under ultra-high mobility, and the selected spectral efficiency strikes a balance between throughput and covertness.  

\subsection{Simulation Results}

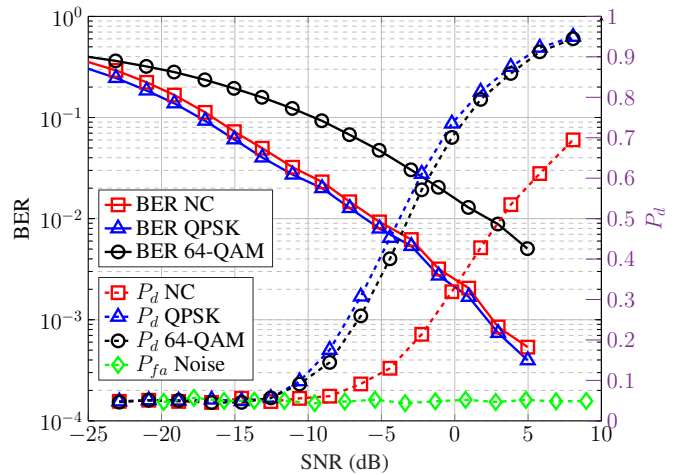
\begin{figure}[t!]
    \centering
%
%
\definecolor{mypurple}{rgb}{0.49,0.18,0.56}

\resizebox{\columnwidth}{!}{

\begin{tikzpicture}

\begin{axis}[%
width=4.521in,
height=3.566in,
at={(0.758in,0.481in)},
scale only axis,
xmin=-25,
xmax=10,
font = \Large,
xlabel style={font=\Large\color{white!15!black}},
xlabel={SNR (dB)},
separate axis lines,
every outer y axis line/.append style={black},
every y tick label/.append style={font=\Large\color{black}},
every y tick/.append style={black},
ymode=log,
ymin=0.0001,
ymax=1,
yminorticks=true,
ylabel={BER},
axis background/.style={fill=white},
xmajorgrids,
ymajorgrids,
yminorgrids,
minor grid style = dashed,
legend style={at={(0.02,0.38)}, anchor=south west, legend cell align=left, align=left, draw=white!15!black}
]
\addplot [color=red, line width=1.5pt, mark size=4pt, mark=square, mark options={solid, red}]
  table[row sep=crcr]{%
-25.0458404701051	0.355432222222222\\
-23.1079938074025	0.292420277777778\\
-21.0164604855271	0.222987222222222\\
-19.1246387186904	0.166766111111111\\
-17.0230366707537	0.11245\\
-15.0350958528381	0.072518611111111\\
-13.1296658609896	0.0496666666666667\\
-11.0598766458951	0.0322186111111111\\
-9.04788427820295	0.0230930555555556\\
-7.16150870515642	0.0146858333333333\\
-5.13301366511671	0.00927888888888889\\
-2.96483990677432	0.00621222222222222\\
-1.09882272547661	0.00318166666666667\\
0.95034533366151	0.002055\\
2.95928652354799	0.0008475\\
4.97810381761434	0.000535833333333333\\
};
\addlegendentry{BER NC}

\addplot [color=blue, line width=1.5pt, mark size=5pt, mark=triangle, mark options={solid, blue}]
  table[row sep=crcr]{%
-25.0456119072228	0.305205555555556\\
-23.1077904450346	0.245819444444444\\
-21.0164562529536	0.184923611111111\\
-19.1248363724228	0.138001944444444\\
-17.0226748287639	0.0929086111111109\\
-15.0351113427816	0.0607597222222221\\
-13.1294748970669	0.040225\\
-11.059550571528	0.0272513888888889\\
-9.04768320534583	0.0200036111111111\\
-7.16149951274469	0.0127172222222222\\
-5.13259675891879	0.00797722222222222\\
-2.96472404718236	0.00535777777777778\\
-1.09822898415493	0.002735\\
0.950287695169017	0.00168361111111111\\
2.95893259407709	0.00074\\
4.97794314038842	0.000396944444444444\\
};
\addlegendentry{BER QPSK}

\addplot [color=black, line width=1.5pt, mark size=4.0pt, mark=o, mark options={solid, black}]
  table[row sep=crcr]{%
-25.0686998621751	0.399553333333333\\
-23.1312581179695	0.363126666666667\\
-21.0394055882051	0.320320555555556\\
-19.1474750603081	0.281730833333333\\
-17.045367934757	0.23611\\
-15.0580489316474	0.194273611111111\\
-13.1528032614478	0.158523333333333\\
-11.0830829025955	0.123007777777778\\
-9.07079105316147	0.0929241666666666\\
-7.18420675357656	0.0676808333333333\\
-5.15593181916238	0.0473994444444444\\
-2.98724903701859	0.0304091666666668\\
-1.12122312780869	0.0204069444444445\\
0.927538414799469	0.0129036111111111\\
2.93607773382579	0.00884388888888889\\
4.95500008972611	0.00505138888888889\\
};
\addlegendentry{BER 64-QAM}

\end{axis}

\begin{axis}[
    width=4.521in,
    height=3.566in,
    at={(0.758in,0.481in)},
    scale only axis,
    font = \Large,
    xmin=-25,
    xmax=9,
    axis y line=right,
    axis x line=none,
    ylabel={$P_d$},
    ylabel style={font=\Large\color{mypurple}},
    ymin=0,
    ymax=1,
    ytick={0,0.1,0.2,0.3,0.4,0.5,0.6,0.7,0.8,0.9,1},
    ytick style={color=mypurple},
    yticklabel style={color=mypurple},
    major tick length = 10pt,
    tick align = inside,
    legend style={at={(0.02,0.1)}, anchor=south west, legend cell align=left, align=left, draw=white!15!black}
]

\addplot [color=red, dashed, line width=1.5pt, mark size=4pt, mark=square, mark options={solid, red}]
  table[row sep=crcr]{%
-22.9444930936234	0.04825\\
-20.9901836329371	0.051\\
-19.0102588692009	0.048\\
-16.8258381027922	0.0455\\
-14.8379742650795	0.056\\
-12.8998961946685	0.04775\\
-10.9991771475578	0.0555\\
-9.00260283697799	0.061\\
-6.94008783271801	0.09125\\
-5.00752222094163	0.13\\
-2.90248920555209	0.214\\
-0.907003550040939	0.31925\\
1.00125981299794	0.428\\
2.99967978110473	0.5345\\
4.91732720805967	0.6115\\
7.12483781019859	0.69475\\
};
\addlegendentry{$P_{d}$ NC}

\addplot [color=blue, dashed, line width=1.5pt, mark size=5pt, mark=triangle, mark options={solid, blue}]
  table[row sep=crcr]{%
-22.9445327107157	0.048\\
-20.9896706119688	0.05075\\
-19.0104592155545	0.05\\
-16.8260714598414	0.0525\\
-14.8385896910383	0.04875\\
-12.8995578520112	0.05625\\
-10.9988480180289	0.09775\\
-9.00279545250967	0.1755\\
-6.94049175549328	0.30725\\
-5.00740373733617	0.45225\\
-2.90234409286599	0.61075\\
-0.907328794081169	0.735\\
1.00109851313818	0.81375\\
2.99947260517518	0.875\\
4.91686968291936	0.9225\\
7.12477494140518	0.9495\\
};
\addlegendentry{$P_{d}$ QPSK}

\addplot [color=black, dashed, line width=1.5pt, mark size=4.0pt, mark=o, mark options={solid, black}]
  table[row sep=crcr]{%
-22.9454501486023	0.04625\\
-20.9902956512131	0.051\\
-19.0109294597652	0.0505\\
-16.8261972010068	0.04525\\
-14.8391512481088	0.04575\\
-12.9007649574322	0.05675\\
-10.9999375226774	0.091\\
-9.00342687037615	0.1445\\
-6.94103702291257	0.2595\\
-5.00811059705206	0.40075\\
-2.90329125126025	0.57125\\
-0.907445797677694	0.70075\\
1.00033871988125	0.79475\\
2.99882423483235	0.859\\
4.91731007679503	0.9125\\
7.12446263750308	0.94475\\
};
\addlegendentry{$P_{d}$ 64-QAM}

\addplot [color=green, dashed, line width=1.5pt, mark size=5pt, mark=diamond, mark options={solid, green}]
  table[row sep=crcr]{%
-20	0.04675\\
-18	0.05675\\
-16	0.04925\\
-14	0.04975\\
-12	0.0495\\
-10	0.04375\\
-8	0.04775\\
-6	0.05225\\
-4	0.04375\\
-2	0.04825\\
0	0.052\\
2	0.04675\\
4	0.05175\\
6	0.04825\\
8	0.049\\
};
\addlegendentry{$P_{fa}$  Noise}

\end{axis}

\end{tikzpicture}%

}
    \vspace{-.25cm}
    \caption{Simulation results: BER and probability of detection $P_d$ for a probability of false alarm $P_{fa} = 0.05$ as a function of the SNR.}
    \label{fig:simulacion}
\end{figure}

Each simulation represents the transmission of one noncoherent (denoted as NC in the figures), one coherent QPSK, and one coherent 64-QAM frame between Alice and Bob, along with a detection attempt by Willie per frame transmission. Even if Willie is able to operate at the same central frequency and sampling rate as Alice, there will be a frequency offset due to the mismatch between the phases of their clocks. Therefore, we introduced a frequency offset between Alice and Willie. The probability of false alarm, obtained when Alice does not transmit and Willie acquires AWGN, has also been estimated. The results obtained by averaging 4000 Monte Carlo simulations for each SNR point are shown in Fig. \ref{fig:simulacion}, which depicts simultaneously the BER on the left side and the probability of detection $P_d$ for a fixed probability of false alarm $P_{fa} =0.05$, on the right side. The proposed noncoherent LPD system provides slightly lower performance than the QPSK system in terms of BER, but because the Grassmannian signals are much more noise-like, its covertness is much higher than that of coherent schemes. As mentioned in \cite{yan_low_2019}, low-order modulations tend to perform better in terms of BER at low SNR regimes. Nevertheless, higher-order modulations tend to better approximate to a Gaussian distribution due to the Central Limit Theorem. The results show that the Gaussianity of a higher-order constellation (64-QAM) is greater than that of a lower-order one (QPSK), but the frequency offset reduces this gap. 

\subsection{Experimental Results}

For the over-the-air (OTA) experimental measurements, we used 3 USRP B210 transceivers controlled from a central computer. Alice's and Bob's USRPs use a common reference signal provided by a reference clock (National Instruments CDA-2990). In an outdoors scenario, this synchronization could be achieved using a GPS Disciplined Oscillator (GPSDO).

\begin{figure}[t!]
    \centering
\includegraphics[width=.75\columnwidth]{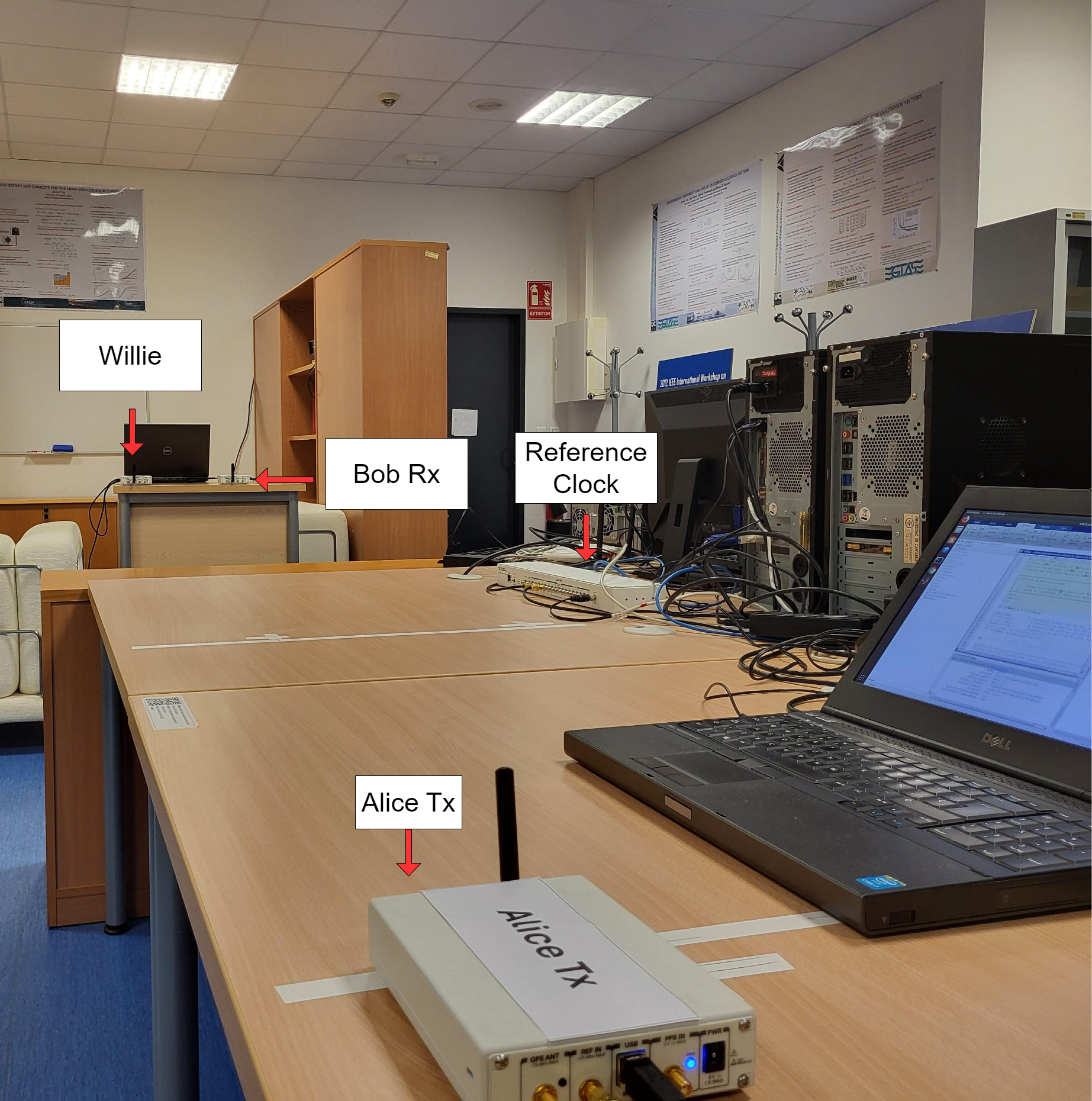}
     \caption{Experimental setup.}
	\label{fig:setup}
\end{figure}

Bob's and Willie's receivers have been placed close to each other, in a clear line-of-sight (LOS) situation towards  Alice, as shown in Fig \ref{fig:setup}, which displays an image of the setup used. It is assumed that all three USRP nodes, Alice, Bob, and Willie, operate at the same central frequency and sampling rate. The experiment was conducted in the 2.5 GHz industrial, scientific, and medical (ISM) band with a sampling frequency of $f_{s}$ = 10 Msamples per second. 

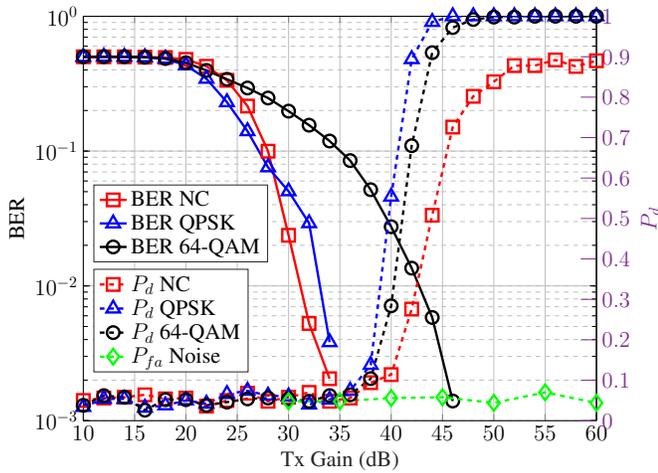
\begin{figure}[t!]
    \centering
%
%
\definecolor{mypurple}{rgb}{0.49,0.18,0.56}

\resizebox{\columnwidth}{!}{

\begin{tikzpicture}

\begin{axis}[%
width=4.521in,
height=3.566in,
at={(0.758in,0.481in)},
scale only axis,
font = \Large,
xmin=10,
xmax=60,
xlabel style={font=\Large\color{white!15!black}},
xlabel={Tx Gain (dB)},
separate axis lines,
every outer y axis line/.append style={black},
every y tick label/.append style={font=\Large\color{black}},
every y tick/.append style={black},
ymode=log,
ymin=0.001,
ymax=1,
yminorticks=true,
ylabel={BER},
axis background/.style={fill=white},
xmajorgrids,
ymajorgrids,
yminorgrids,
minor grid style = dashed,
legend style={at={(0.02,0.392)}, anchor=south west, legend cell align=left, align=left, draw=white!15!black}
]
\addplot [color=red, line width=1.5pt, mark size=4pt, mark=square, mark options={solid, red}]
  table[row sep=crcr]{%
10	0.500542222222223\\
12	0.500491111111111\\
14	0.50032\\
16	0.497353333333333\\
18	0.494379999999999\\
20	0.47892\\
22	0.425817777777778\\
24	0.334933333333334\\
26	0.215673333333334\\
28	0.0998177777777778\\
30	0.0237244444444445\\
32	0.00526444444444444\\
34	0.00204888888888889\\
};
\addlegendentry{BER NC}

\addplot [color=blue, line width=1.5pt, mark size=5pt, mark=triangle, mark options={solid, blue}]
  table[row sep=crcr]{%
10	0.498571111111111\\
12	0.501237777777778\\
14	0.50008\\
16	0.496513333333333\\
18	0.484891111111111\\
20	0.432371111111111\\
22	0.343911111111111\\
24	0.230448888888889\\
26	0.1405\\
28	0.0758133333333333\\
30	0.05034\\
32	0.0291844444444445\\
34	0.00382444444444442\\
};
\addlegendentry{BER QPSK}

\addplot [color=black, line width=1.5pt, mark size=4.0pt, mark=o, mark options={solid, black}]
  table[row sep=crcr]{%
10	0.50142\\
12	0.501233333333334\\
14	0.501373333333334\\
16	0.496484444444445\\
18	0.486073333333333\\
20	0.452064444444444\\
22	0.396651111111111\\
24	0.340764444444444\\
26	0.293706666666667\\
28	0.247493333333333\\
30	0.197777777777778\\
32	0.155644444444444\\
34	0.118742222222222\\
36	0.0850155555555555\\
38	0.0517044444444446\\
40	0.0274733333333333\\
42	0.0135377777777778\\
44	0.00583111111111111\\
46	0.00139999999999999\\
};
\addlegendentry{BER 64-QAM}

\end{axis}

\begin{axis}[
    width=4.521in,
    height=3.566in,
    at={(0.758in,0.481in)},
    scale only axis,
    font = \Large,
    xmin=10,
    xmax=60,
    axis y line=right,
    axis x line=none,
    ylabel={$P_d$},
    ylabel style={font=\Large\color{mypurple}},
    ymin=0,
    ymax=1,
    ytick={0,0.1,0.2,0.3,0.4,0.5,0.6,0.7,0.8,0.9,1},
    ytick style={color=mypurple},
    yticklabel style={color=mypurple},
    major tick length = 10pt,
    tick align = inside,
    legend style={at={(0.02,0.12)}, anchor=south west, legend cell align=left, align=left, draw=white!15!black}
]

\addplot [color=red, dashed, line width=1.5pt, mark size=4pt, mark=square, mark options={solid, red}]
  table[row sep=crcr]{%
10	0.05\\
12	0.056\\
14	0.058\\
16	0.064\\
18	0.054\\
20	0.056\\
22	0.036\\
24	0.05\\
26	0.068\\
28	0.048\\
30	0.058\\
32	0.07\\
34	0.048\\
36	0.056\\
38	0.094\\
40	0.114\\
42	0.276\\
44	0.508\\
46	0.726\\
48	0.802\\
50	0.838\\
52	0.878\\
54	0.878\\
56	0.892\\
58	0.876\\
60	0.89\\
};
\addlegendentry{$P_{d}$ NC}

\addplot [color=blue, dashed, line width=1.5pt, mark size=5pt, mark=triangle, mark options={solid, blue}]
  table[row sep=crcr]{%
10	0.034\\
12	0.054\\
14	0.054\\
16	0.032\\
18	0.038\\
20	0.05\\
22	0.042\\
24	0.066\\
26	0.074\\
28	0.062\\
30	0.06\\
32	0.04\\
34	0.052\\
36	0.074\\
38	0.138\\
40	0.554\\
42	0.894\\
44	0.986\\
46	1\\
48	1\\
50	1\\
52	1\\
54	1\\
56	1\\
58	1\\
60	1\\
};
\addlegendentry{$P_{d}$ QPSK}

\addplot [color=black, dashed, line width=1.5pt, mark size=4.0pt, mark=o, mark options={solid, black}]
  table[row sep=crcr]{%
10	0.038\\
12	0.062\\
14	0.058\\
16	0.026\\
18	0.052\\
20	0.052\\
22	0.038\\
24	0.046\\
26	0.054\\
28	0.056\\
30	0.054\\
32	0.048\\
34	0.062\\
36	0.064\\
38	0.104\\
40	0.284\\
42	0.68\\
44	0.91\\
46	0.972\\
48	0.992\\
50	0.998\\
52	0.998\\
54	1\\
56	1\\
58	1\\
60	1\\
};
\addlegendentry{$P_{d}$ 64-QAM}

\addplot [color=green, dashed, line width=1.5pt, mark size=5pt, mark=diamond, mark options={solid, green}]
  table[row sep=crcr]{%
30	0.048\\
35	0.048\\
40	0.056\\
45	0.058\\
50	0.044\\
55	0.07\\
60	0.044\\
};
\addlegendentry{$P_{fa}$ Noise}

\end{axis}

\end{tikzpicture}%

}
    \vspace{-.5cm}
    \caption{Experimental results: BER and probability of detection for a probability of false alarm $P_{fa} = 0.05$ as a function of the SNR.}
    \label{fig:setupResults}
\end{figure}

For the experimental results, the same methodology used for the simulations has been followed, with the exception that instead of defining the operating point in terms of the SNR, it has been determined using Alice's transmitter gain. Consequently, a sweep in transmitter gain has been conducted. This approach was chosen due to the practical difficulties in estimating the noise at Bob's receiver. The results obtained in the experiment are shown in Fig. \ref{fig:setupResults}, where we can see that the slope of the curves has changed compared to those obtained by simulation (Fig \ref{fig:simulacion}). This is because the channel is now similar to an AWGN channel, since the environment in which the measurements were conducted is static (low time selectivity) and the channel is flat in frequency due to its strong line-of-sight (LOS) component. In terms of detectability, we observe how the experimental results match those obtained through simulation, although the gap between the noncoherent and coherent schemes has narrowed slightly. This is because the simulation assumes perfect synchronization and ideal hardware components, whereas in OTA experiments the time synchronization is not perfect and hardware non-idealities affect the results.

\section{Conclusion}

In this paper, we have proposed a noncoherent communications system for LPD communications based on DSSS and Grassmannian signaling. The performance of the proposed system has been evaluated both by simulations and by conducting OTA experiments with three USRP nodes. Compared to coherent schemes that use QPSK or QAM modulation formats and need pilots to perform channel estimation, the proposed noncoherent scheme demonstrated promising performance in terms of BER at low SNR regimes while achieving a distribution closer to Gaussian at Willie's detector, thereby providing better covertness. Future work may explore the extension of this approach to multiple-input multiple-output (MIMO) systems and the integration of additional PHY-layer security techniques to further enhance covertness and performance \cite{PhySec_ICASSP2014}.

\balance
\bibliographystyle{IEEEtran}
\bibliography{LPD_paper}

\end{document}